\documentclass[12pt]{article}   
\usepackage{amsmath, amsthm, amscd, amsfonts, amssymb, graphicx, color}
\usepackage[utf8]{inputenc}
\usepackage{cite}
\usepackage[T1]{fontenc}
\usepackage{graphicx}
\usepackage{epstopdf} 
\usepackage{color}
\input{epsf}
\usepackage{lmodern}
\usepackage{wrapfig}
\usepackage{amsmath}
\newcommand{\be}{\begin{equation}}
\newcommand{\ee}{\end{equation}}
\newcommand{\bear}{\begin{eqnarray}}
\newcommand{\ear}{\end{eqnarray}}

\begin{document}
\title{Bound states in a semi-infinite square potential well}
\author{Nivaldo A. Lemos  \\
\small
{\it Instituto de F\'{\i}sica - Universidade Federal Fluminense}\\
\small
{\it Av. Litor\^anea s/n, Boa Viagem, 
24210-340, Niter\'oi - RJ, Brasil}\\
\small
{\it  nivaldolemos@id.uff.br}}

\date{\today}

\maketitle

\begin{abstract}

The finite square potential well is a  staple problem in introductory quantum mechanics.  There is an extensive literature on the determination of the allowed energies, which requires the solution of a transcendental equation  by numerical, graphical or approximate analytic methods. Here we investigate the less explored problem of a particle in a semi-infinite potential well. The energy eigenvalues, which are also determined by a transcendental equation,  are found by a standard graphical method, and a simple rule that yields the number of stationary states is provided. Next a simplification of the aforementioned transcendental equation is attempted. During the process pitfalls are encountered and a purportedly simpler graphical treatment of the problem given in the solutions manual to a fine textbook is shown to be flawed. A more careful analysis brings forth the correct simplification, which is shown to be particularly suitable for finding highly accurate approximations to the energy levels. Finally, a class of exact solutions is produced,  
the associated normalized eigenfunctions are constructed and the probability of finding the particle inside the well is computed.

\end{abstract}

\maketitle

\section{Introduction}\label{Intro}

 In modern physics or introductory quantum mechanics courses the time-independent Schr\"odinger equation is first applied to one-dimensional  physical systems such as a particle in a
potential well \cite{Serway,Schiff}. These examples illustrate the quantization of energy and how the allowed energies arise 
from the imposition of appropriate boundary conditions on the wave function.
The case of the infinite square well --- or particle in a box --- can be easily solved exactly, but for the less unrealistic finite
potential well the  allowed energies arise as solutions of a transcendental
equation, which can be found only numerically \cite{Murphy},  graphically  \cite{Pitkanen,Cantrell,Guest,Mallow} or by analytic
approximation methods \cite{Garrett,Searles,Barker,Aronstein,Bloch,Griffiths,Naqvi,Lima}.
Given its apparent inexhaustibility, the recent appearence of a guide to the  large literature on the finite square well problem is a good thing \cite{Reed}.
The finite square well has a wide range of practical applications,  from   the theory of alpha decay (square  well coupled to Coulomb barrier) in nuclear physics  \cite{Martin} to lasers and condensed matter physics \cite{Wang}.

Here we turn our attention to the less studied problem of the semi-infinite square  well \cite{Griffiths,Zettili,Lopez}. Scattering states are excluded from the discussion, the focus is on bound states. This is  usually a textbook problem \cite{Serway2,Zettili} but it also finds application in condensed matter physics \cite{Li} and in  modelling the nuclear potential \cite{Martin}.  Similarly to the finite  well, the determination of the allowed energy levels for the semi-infinite well  requires the solution of a transcendental equation. 

In Section \ref{Semi-infinite} the problem is stated and the transcendental equation whose solutions furnish the possible energies is derived. 
In Section \ref{graphical} we discuss the standard graphical solution and supply an exact and practical way to determine  the number of stationary states. Next, in Section \ref{Simplifying}, we address some attempts at simplification of the transcendental equation that yields the allowed energies. In this process,  pitfalls are recognized and it is pointed out that the reasoning pursued in the solutions manual to Ref. \cite{Serway} does not pass muster. The correct simplification is found and, in Section \ref{Newton-method}, it is shown to be especially apt for getting strikingly accurate approximations to the energy levels.  Finally, in Section \ref{Exact}, a class of exact solutions for the energy is obtained. The associated wave functions are explicitly normalized and the probability of finding the particle inside the well is calculated.  
Section \ref{Conclusion} is devoted to a few  remarks of a general nature.

\section{Semi-infinite square well in quantum mechanics}\label{Semi-infinite}

\begin{wrapfigure}{r}{0.35\textwidth}
\vspace{-25pt}
\begin{center}
\includegraphics[width=0.35\textwidth]{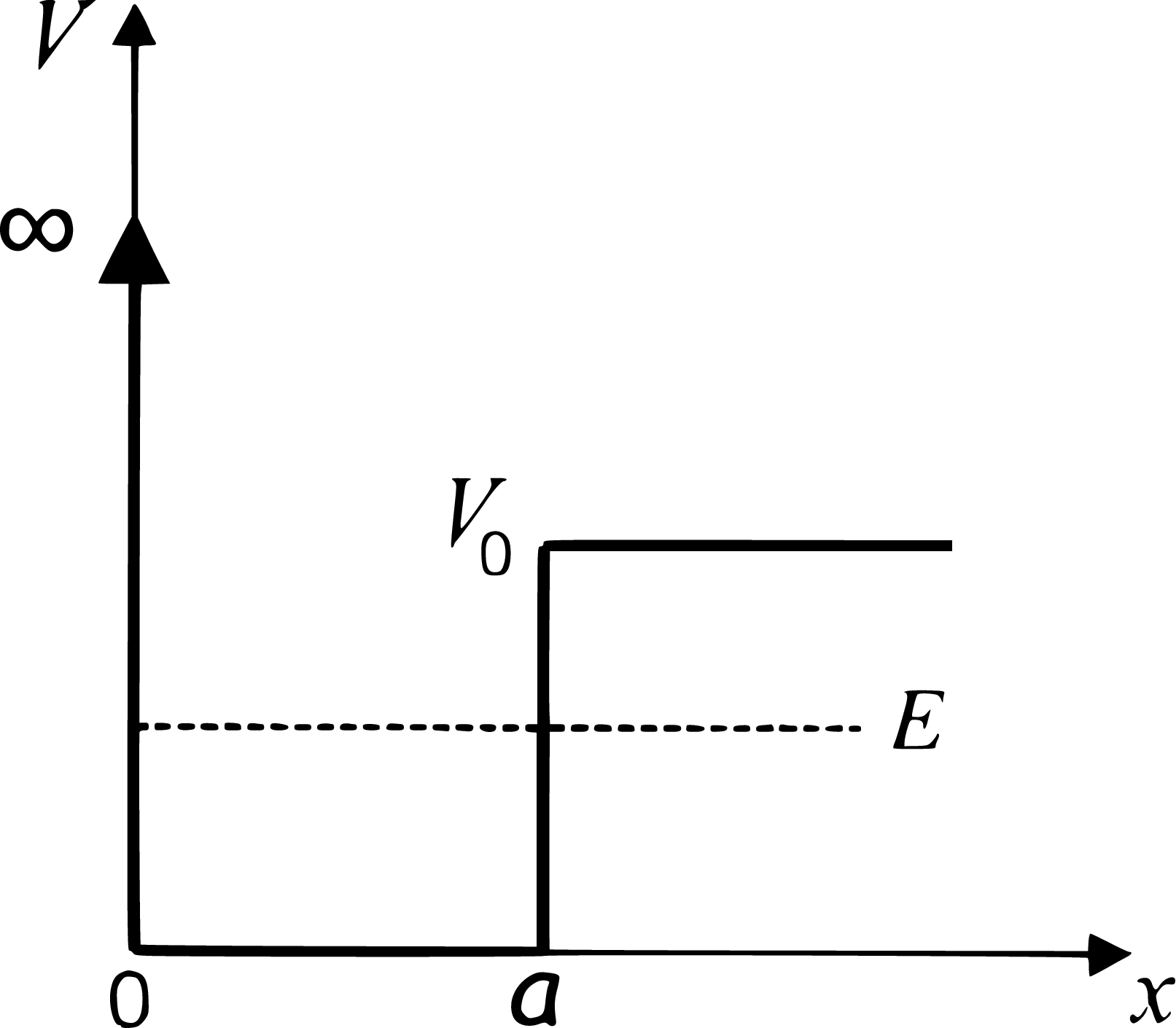}
\vspace{-20pt}
\end{center}
\vspace{-10pt}
\caption{Semi-infinite potential energy square well. The particle is excluded from the region $x < 0$ by an impenetrable wall of infinitely high potential energy. }
\label{Figure-potential}
\end{wrapfigure}

Consider a particle of mass $m$ subject to the potential energy defined by
 \begin{equation}
\label{potential}
 V(x) = \left\{ \begin{array}{cl}
                    \infty & \mbox{if $x<0$} \\
                       0  & \mbox{if $0 \leq x \leq a$}\\
                      V_0 & \mbox{if $x>a$}
                  \end{array} \right..
\end{equation}
 This potential energy is depicted in Fig. \ref{Figure-potential}. The  positive constants $a$ and $V_0$ characterize the well's width and depth, respectively. The particle is kept away from the region $x<0$ by an infinitely high potential barrier.

There can be definite energy bound states, described by square integrable wave functions, only if the energy $E$ is such that $0< E < V_0$, which is assumed from now on.

The one-dimensional time-independent Schr\"odinger equation reads
\begin{equation}
\label{Schrodinger}
 \frac{d^2 \psi}{dx^2} + \frac{2m(E-V)}{\hbar^2} \psi = 0.
\end{equation}
Since the infinite potential barrier demands $\psi (x)=0$ for $x<0$, the boundary conditions on the physically acceptable solutions to equation \eqref{Schrodinger} are
\begin{equation}
\label{Boundary}
 \psi (0) =0; \qquad \psi, \psi^{\prime} \,\, \mbox{continuous at}\,\, x=a.
\end{equation}

Because the potential energy is  discontinuous at $x=a$, the time-independent Schr\"odinger equation must be set up separately  for each of the regions  $0 \leq x \leq a$ and $x> a$.

$\blacksquare\,\,$ Inside the wall $(0 \leq x \leq a)$. Taking into account that $V=0$, equation \eqref{Schrodinger} becomes
\begin{equation}
\label{Schrodinger-inside}
 \frac{d^2 \psi}{dx^2} + k^2 \psi = 0
\end{equation}
where
\begin{equation}
\label{k}
 k = \frac{\sqrt{2mE}}{\hbar}.
\end{equation}
The general solution to equation \eqref{Schrodinger-inside} is
\begin{equation}
\label{psi-inside-general}
 \psi (x) = A \sin kx + C \cos kx,
\end{equation}
where $A$ and $C$ are arbitrary constants. The condition $\psi (0) =0$ yields $C=0$. Therefore,
\begin{equation}
\label{psi-inside}
 \psi (x) = A \sin kx \qquad\qquad (0 \leq x \leq a).
\end{equation}

\medskip

$\blacksquare\,\,$ Outside the wall $(x > a)$. Given that $V=V_0>E$, equation \eqref{Schrodinger} takes the form
\begin{equation}
\label{Schrodinger-outside}
 \frac{d^2 \psi}{dx^2} - {\tilde k}^2 \psi = 0
\end{equation}
where
\begin{equation}
\label{k-tilde}
 {\tilde k} = \frac{\sqrt{2m(V_0 -E)}}{\hbar}.
\end{equation}
The general solution to equation \eqref{Schrodinger-outside} is
\begin{equation}
\label{psi-outside-general}
 \psi (x) = B e^{-{\tilde k}x} + De^{{\tilde k}x},
\end{equation}
where $B$ and $D$ are arbitrary constants. The positive exponential must be discarded because it grows without bound as $x \to \infty$ and would give rise to a non-normalizable wave function. Thus, one must set $D=0$, which yields
\begin{equation}
\label{psi-outside}
 \psi (x) = B e^{-{\tilde k}x} \qquad\qquad (x > a).
\end{equation}

\bigskip

 Continuity of $\psi$ and $\psi^{\prime}$ at $x=a$ leads to
\begin{eqnarray}
\label{boundary-conditions-x=a-1}
 A \sin ka & = & B e^{-{\tilde k}a},\\
\label{boundary-conditions-x=a-2}
A k\cos ka & = & -B {\tilde k} e^{-{\tilde k}a}.
\end{eqnarray}
If $A=0$, equation \eqref{boundary-conditions-x=a-1} implies that $B=0$. This is unacceptable because it would give $\psi =0$, meaning that the particle does not exist --- it is nowhere to be found. Since $ \sin ka$ and $ \cos ka$ cannot be zero at the same time, it cannot be the case that $B=0$ because this would imply $A=0$. Finally, $ \sin ka$ and $ \cos ka$ must be both nonvanishing, otherwise one would have $B=0$, which has just been seen to be inadmissible. Therefore, both sides of   equations \eqref{boundary-conditions-x=a-1} and \eqref{boundary-conditions-x=a-2}
are nonzero, and one can safely divide the second equation by the first to get
\begin{equation}
\label{boundary-conditions-divided}
 {\tilde k} = - k \cot ka .  
\end{equation} 
 Since both $k$ and $\tilde k$ depend on $E$, this transcendental equation determines the allowed energies.

It is worth noting that this is the same transcendental equation that  gives the energies associated with the odd solutions to the time-independent Schr\"odinger equation  for the finite square well when its middle point is chosen as the origin of the $x$-axis \cite{Schiff}.

\section{Graphical determination of the energy levels}\label{graphical}

Except for certain especial values of  $V_0$, exact solutions to the transcendental equation \eqref{boundary-conditions-divided} are not known. 
Therefore, one must resort to  numerical, graphical or approximate algebraic methods for solving it.
From \eqref{k} and \eqref{k-tilde} it follows that 
\begin{equation}
\label{k-ktilde-condition}
 k^2 +{\tilde k}^2 = \frac{2mV_0}{\hbar^2} .  
\end{equation} 
It is useful to introduce the positive dimensionless quantities $z, {\tilde z}, z_0$  defined by
\begin{equation}
\label{z-ztilde-z-zero}
z = ka, \qquad {\tilde z} = {\tilde k}a, \qquad z_0 = \sqrt{\frac{2mV_0a^2}{\hbar^2}},  
\end{equation} 
in terms of which equation \eqref{k-ktilde-condition} becomes
\begin{equation}
\label{z-ztilde-condition}
 z^2 +{\tilde z}^2 = z_0^2,
\end{equation} 
which represents a circle of radius $z_0$ with center at the origin of the $z{\tilde z}$-plane.

So, equation \eqref{boundary-conditions-divided} can be written as
\begin{equation}
\label{transcendental-equation-exact}
 \sqrt{z_0^2 - z^2} = - z \cot z .  
\end{equation} 
Graphically, the allowed energies are determined from the $z$-values corresponding to the intersections of the first quadrant of the origin-centered circle of radius $z_0$  with the graph of the function ${\tilde z} = - z \cot z$ in the
 region $z>0$ of the $z{\tilde z}$-plane. From equations \eqref{k} and \eqref{z-ztilde-z-zero}, to each $z_n$ that satisfies equation \eqref{transcendental-equation-exact} there corresponds the allowed energy
\begin{equation}
\label{allowed-energy}
E_n = \frac{\hbar^2 z_n^2}{2ma^2}.  
\end{equation} 

Note that $-z\cot z \leq 0$ for $z \in [0,\pi/2]$. Since necessarily $z < z_0$ and the left-hand side of \eqref{transcendental-equation-exact} is positive, that equation has no solutions if 
\begin{equation}
\label{bound-z-zero}
z_0 \leq  \frac{\pi}{2}.
\end{equation}
According to \eqref{z-ztilde-z-zero}, this means  that there are no bound states as long as
\begin{equation}
\label{condition-Vzero}
V_0 a^2 \leq \frac{\pi^2\hbar^2}{8m}.
\end{equation} 
If the well is too shallow or too narrow there are no bound states. This distiguishes the semi-infinite well from the finite one, for which there is always at least one bound state.

\begin{figure}[h!]
\begin{center}
 \includegraphics[width=.45\textwidth]{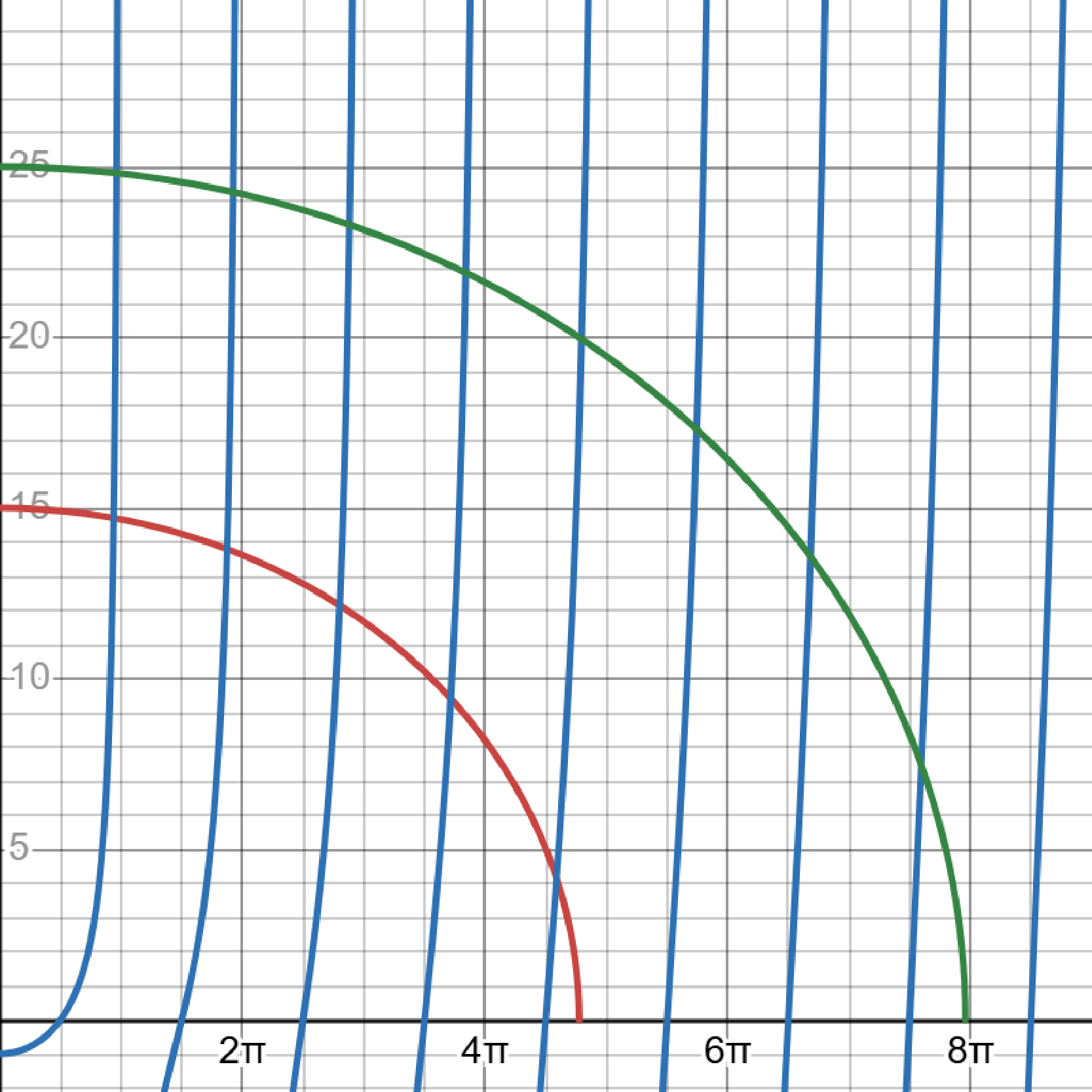}
\caption{Graphs of the function ${\tilde z} = -z \cot z$ and of the first quadrant of the circle $z^2 + {\tilde z}^2 = z_0^2$  for  $z_0=15$ and $z_0=25$. The circular arcs are slightly distorted because the horizontal and vertical scales are  not quite the same. Values of $z$ are  on the horizontal axis while those of $\tilde z$ are on the vertical axis. The value of $z$ for each intersection of the two graphs is a solution to equation \eqref{transcendental-equation-exact}. For $z_0=15$ there are 5 solutions, whereas for $z_0=25$ there are 8 solutions. The number of solutions is equal to the positive integer $N$ that satisfies inequalities \eqref{number-of-solutions}.}
\label{Figure-intersections}
\end{center}
\end{figure}

The graphical solution\footnote{All graphical and numerical computations have been performed at https://www.desmos.com/?lang=en.} to equation \eqref{transcendental-equation-exact} is given in Fig. \ref{Figure-intersections} for $z_0=15$ and $z_0=25$. There are five solutions for $z_0=15$ and eight solutions for $z_0=25$.
The function $z \mapsto  \cot z$ vanishes at $z= (2n-1)\frac{\pi}{2}$ where $n=1,2,3, \ldots$. Figure \ref{Figure-intersections} makes it clear that there is only one intersection of the graphs 
 between two consecutive zeroes of $\cot z$, that is, in each interval $(2n-1)\frac{\pi}{2} < z < (2n+1)\frac{\pi}{2}$. Thus, the number of bound states is given by the  number that counts the last intersection, namely the positive integer  $N$ such that $ (2N-1)\frac{\pi}{2} < z_0 < (2N+1)\frac{\pi}{2}$, which is equivalent to
\begin{equation}
\label{number-of-solutions}
2N-1 < \frac{2z_0}{\pi} < 2N+1.
\end{equation}
Except for the form, these inequalities for determination of the number of bound states appear in \cite{Zettili} and \cite{Lopez}. For $z_0=15$ we have $2z_0/\pi = 9.55$ with $9 <  9.55 < 11$, which gives $N=5$; if  $z_0=25$ we have $2z_0/\pi = 15.9$ with  $15 < 15.9 < 17$, which yields $N=8$. These results are born out by Fig. \ref{Figure-intersections}. Note that if $z_0 \leq \pi/2$ then $2z_0/\pi \leq 1$ and no positive integer $N$ satisfies inequalities \eqref{number-of-solutions}, a confirmation that in this case no bound state exists.\footnote{ If $z_0=m\pi/2$ with  $m>1$ an odd integer, then $z_0$ is a zero of $\cot z$ and the last  intersection occurs at $z=z_0$. This solution must be discarded because $z=z_0$ implies ${\tilde z}=0$, which is unacceptable. In this case, the number $N$ of bound states is still given by \eqref{number-of-solutions} as long as $2z_0/\pi$ is replaced with the even number $m-1$.}

For the record, we give below the numerical values up to five decimal places of the solutions to equation \eqref{transcendental-equation-exact} for the two values of $z_0$ considered in Fig. \ref{Figure-intersections}.
\bigskip\smallskip

\centerline{The five solutions $\, z_1, \ldots ,z_5\,$  for $z_0=15$:}
\medskip
\centerline{$2.94404;\quad  5.88035; \quad  8.79801; \quad  11.67442;\quad    14.41691.$}

\bigskip
\centerline{The eight  solutions $\, z_1, \ldots ,z_8\,$ for $z_0=25$:}
\medskip
\centerline{$3.02048; \quad 6.03920;\quad  9.05419; \quad 12.06285; \quad  15.06139; \quad   18.04326; \quad   20.99429; \quad   23.86449.$}

\section{Simplifying the search for the energy eigenvalues}\label{Simplifying}

The transcendental equation for the energy levels can be simplified as follows. Upon multiplying equation \eqref{boundary-conditions-divided} by the well's width $a$ one gets
\begin{equation}
\label{ztilde-function-z}
{\tilde z} = - z \cot z.
\end{equation}
Combining this with equation \eqref{z-ztilde-condition} and taking into account that $1+\cot^2z=\csc^2z=1/\sin^2z$, one arrives at
\begin{equation}
\label{ztilde-function-z-squared}
z^2 = z_0^2 \sin^2z.
\end{equation}
This seems to lead to
\begin{equation}
\label{ztilde-function-z-no-square}
z = z_0 \sin z,
\end{equation}
which is far simpler than \eqref{transcendental-equation-exact} and gives the energy eigenvalues in terms of the intersections of the straight line through the origin $\displaystyle {\tilde z} = z/z_0$ with the graph of the sine function ${\tilde z} = \sin z$. This is how the problem is treated in \cite{Serway3}. Accordingly, since $z > \sin z$ for all $z>0$, there is no bound state if $ z_0 \leq 1$, which is equivalent to $\displaystyle V_0 a^2 \leq \frac{\hbar^2}{2m}$. Unfortunately, this disagrees with \eqref{condition-Vzero} and leads to the incorrect prediction of bound states that actually do not exist.


\begin{figure}[h!]
\begin{center}
 \includegraphics[width=.5\textwidth]{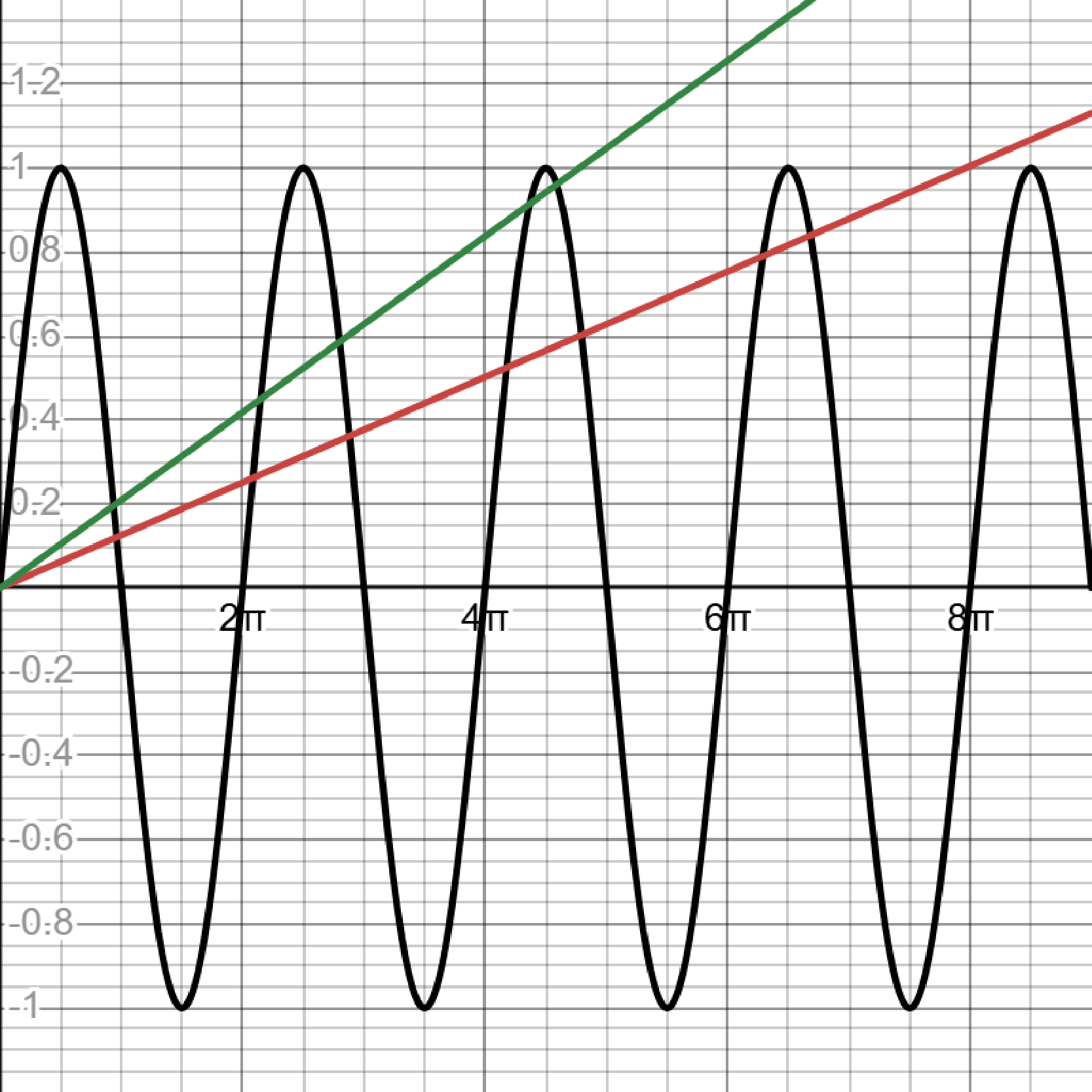}
\caption{Graphical solutions of equation \eqref{ztilde-function-z-no-square} for $z_0=15$ (steeper straight line) and $z_0=25$ (less slanted straight line). }
\label{Figure-intersections-z-sinz-2}
\end{center}
\end{figure}

Figure \ref{Figure-intersections-z-sinz-2} shows the graphical solutions  to equation \eqref{ztilde-function-z-no-square} for $z_0=15$ and  $z_0=25$. For $z_0=15$ there are 5 intersections, the same number shown in Fig. \ref{Figure-intersections}. However, the second and fourth intersections are not solutions to the transcendental equation \eqref{transcendental-equation-exact} because they correspond to values of $z$ for which $\cot z > 0$. Only the first, third and fifth intersections provide true energy levels.  For $z_0=25$ there are 7 intersections, one less than the number of intersections  in Fig. \ref{Figure-intersections}. The second, fourth and sixth intersections are spurious solutions to the transcendental equation \eqref{transcendental-equation-exact}  because they correspond to values of $z$ such that $\cot z > 0$. Only the first, third, fifth and seventh intersections provide true energy levels. It is clear, therefore, that \eqref{ztilde-function-z-no-square} is not the correct equation for the determination of the energy levels, and also it does not give rise to the correct condition  \eqref{condition-Vzero} for the nonexistence of bound states.

\subsection{Another simplification attempt}

Since $z>0$, perhaps the correct consequence to be derived from \eqref{ztilde-function-z-squared} is
\begin{equation}
\label{ztilde-function-z-square-root}
z = z_0 \vert\sin z\vert.
\end{equation}
According to Fig. \ref{Figure-intersections-z-modulus-sinz-and-negative-sinz}	(left),  now there are nine intersections for $z_0=15$, four of which are spurious (the even-numbered ones).  Similarly, there are fifteen  intersections for $z_0=25$, seven of which are spurious (the even-numbered ones). The extraneous solutions are those such that $\cot z >0$. Equation \eqref{ztilde-function-z-square-root} gives the correct solutions and the correct number of solutions if it is supplemented with the condition $\cot z <0$, entailing that $z$ must lie in the second or fourth quadrants. Nevertheless, it is not convenient for a graphical solution because it does not provide a bijective correspondence between intersections and energy levels.

\subsection{Yet another simplification attempt}

Maybe the right thing to do is to take the negative square root of equation \eqref{ztilde-function-z-squared} to obtain
\begin{equation}
\label{ztilde-function-z-negative-square-root}
z = - z_0 \sin z.
\end{equation}

\begin{figure}[!htb]
    \centering
    \begin{minipage}{.5\textwidth}
        \centering
        \includegraphics[width=0.7\linewidth, height=0.3\textheight]{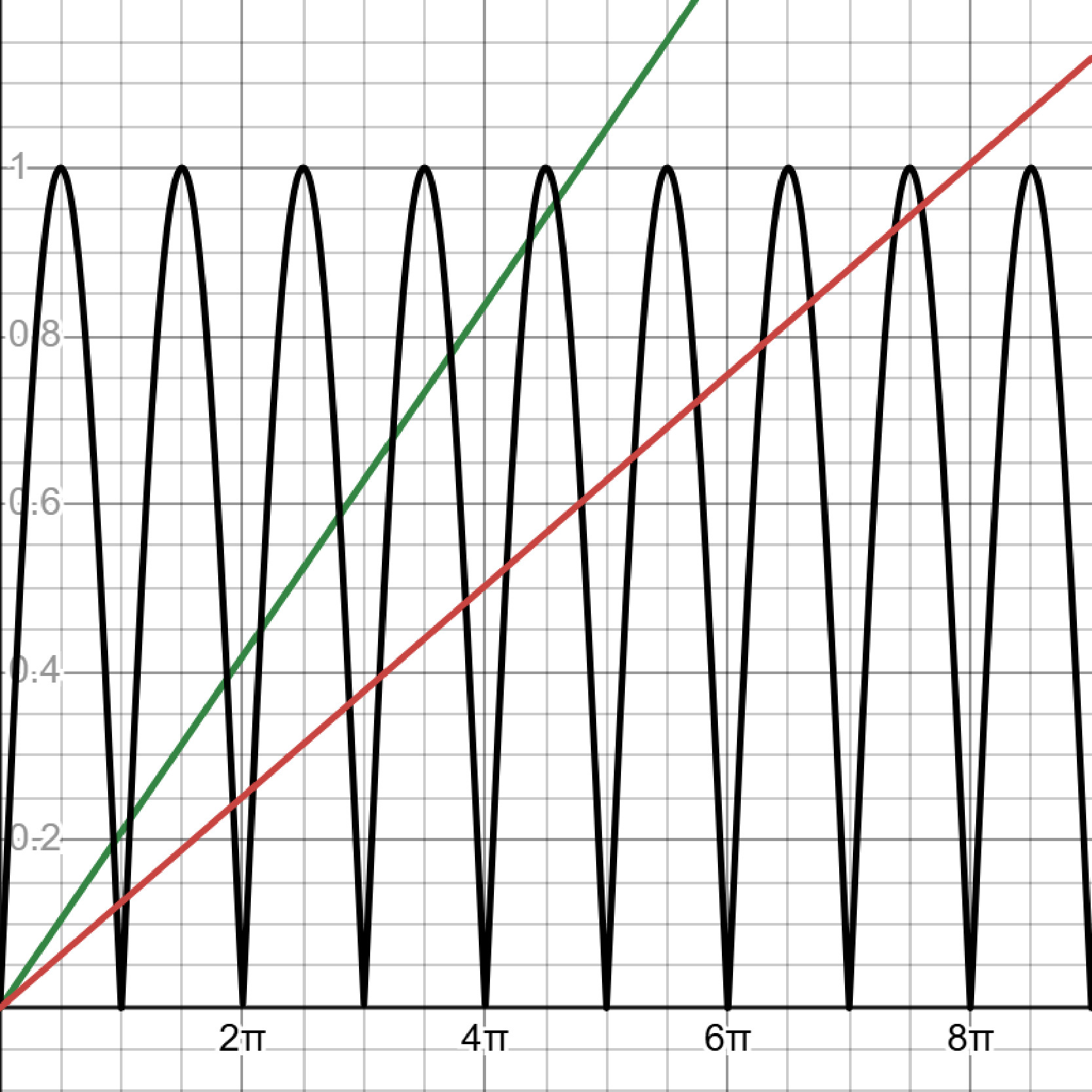}
    \end{minipage}%
    \begin{minipage}{0.5\textwidth}
        \centering
        \includegraphics[width=0.7\linewidth, height=0.3\textheight]{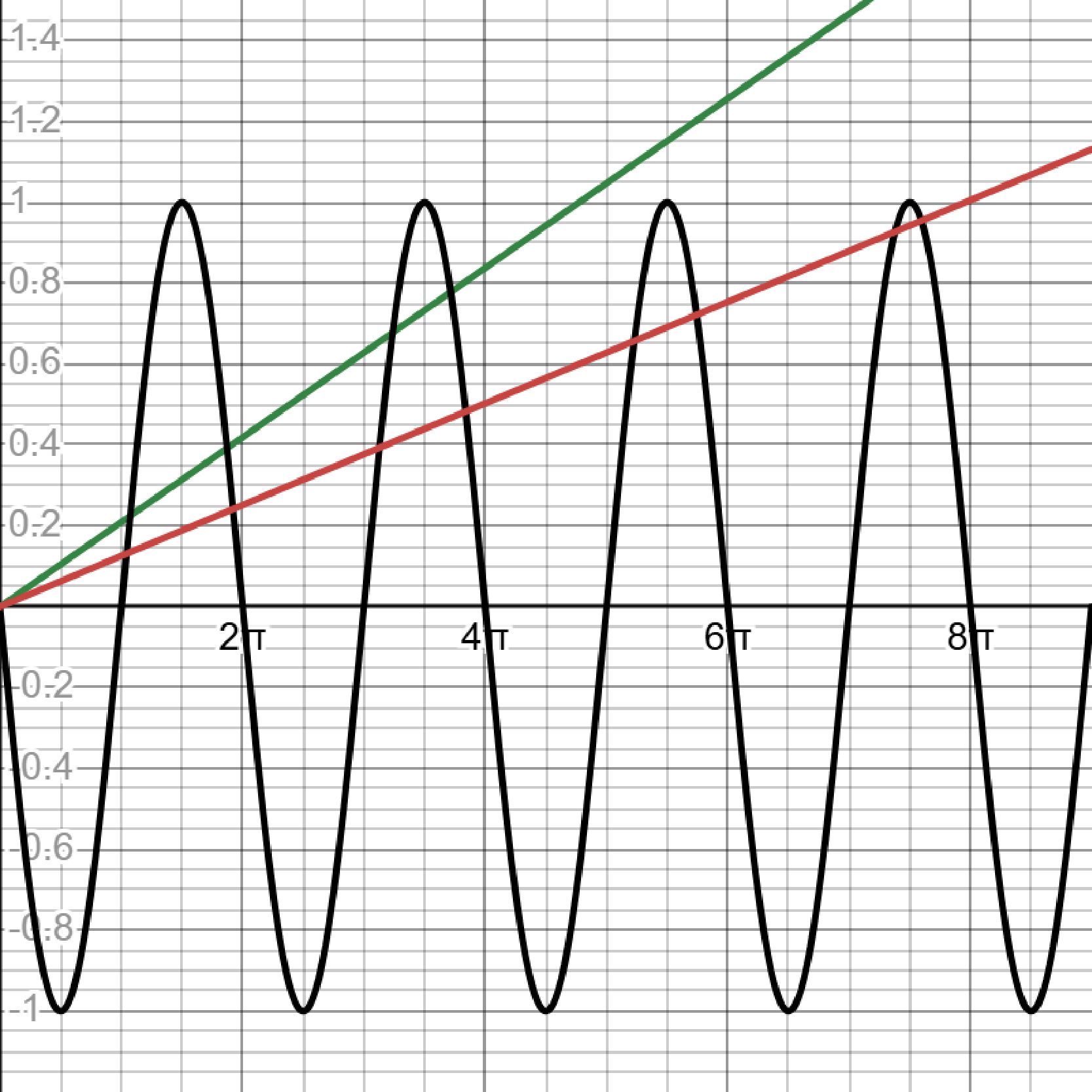}
    \end{minipage}
		\caption{Left: solutions of equation \eqref{ztilde-function-z-square-root} for $z_0=15$  and $z_0=25$. Right: solutions of equation \eqref{ztilde-function-z-negative-square-root} for $z_0=15$ and $z_0=25$. In both cases, the steeper straight line corresponds to $z_0=15$.}
\label{Figure-intersections-z-modulus-sinz-and-negative-sinz}		
\end{figure}

As shown in  Fig. \ref{Figure-intersections-z-modulus-sinz-and-negative-sinz}	(right),  there are only four intersections for $z_0=15$, two of which are spurious (the first and the third).  Similarly, there are eight  intersections for $z_0=25$, half  of which are spurious (the odd-numbered ones). Once again, the extraneous solutions are those such that $\cot z >0$. Equation \eqref{ztilde-function-z-negative-square-root} is incorrect because it does not give the right number of solutions, and even if supplemented with the condition $\cot z <0$ it misses half of the solutions.

\subsection{The correct simplification}

Inserting \eqref{ztilde-function-z-squared} into \eqref{transcendental-equation-exact} and taking into account that $\sqrt{x^2}=\vert x\vert$ one arrives at
\begin{equation}
\label{ztilde-eliminated}
z_0 \vert \cos z\vert  = -  z  \cot z,
\end{equation}
which is equivalent to
\begin{equation}
\label{transcendental-sinz-modulus-cosz}
z = -z_0 \frac{\sin z \cos z}{\vert \cos z\vert}.
\end{equation}
First of all,  note that this equation does not admit solutions if $z_0 < \pi/2$ because both $\sin z > 0$ and $\cos z >0$ for $0 < z < \pi/2$. Thus, it gives rise to the correct condition \eqref{condition-Vzero} that prevents the existence of bound states.  Furthermore, this equation has solutions only if $\cos z$ and $\sin z$ have opposite signs, implying that $\cot z <0$ as required by equation \eqref{transcendental-equation-exact}.

\begin{figure}[h!]
\begin{center}
 \includegraphics[width=.5\textwidth]{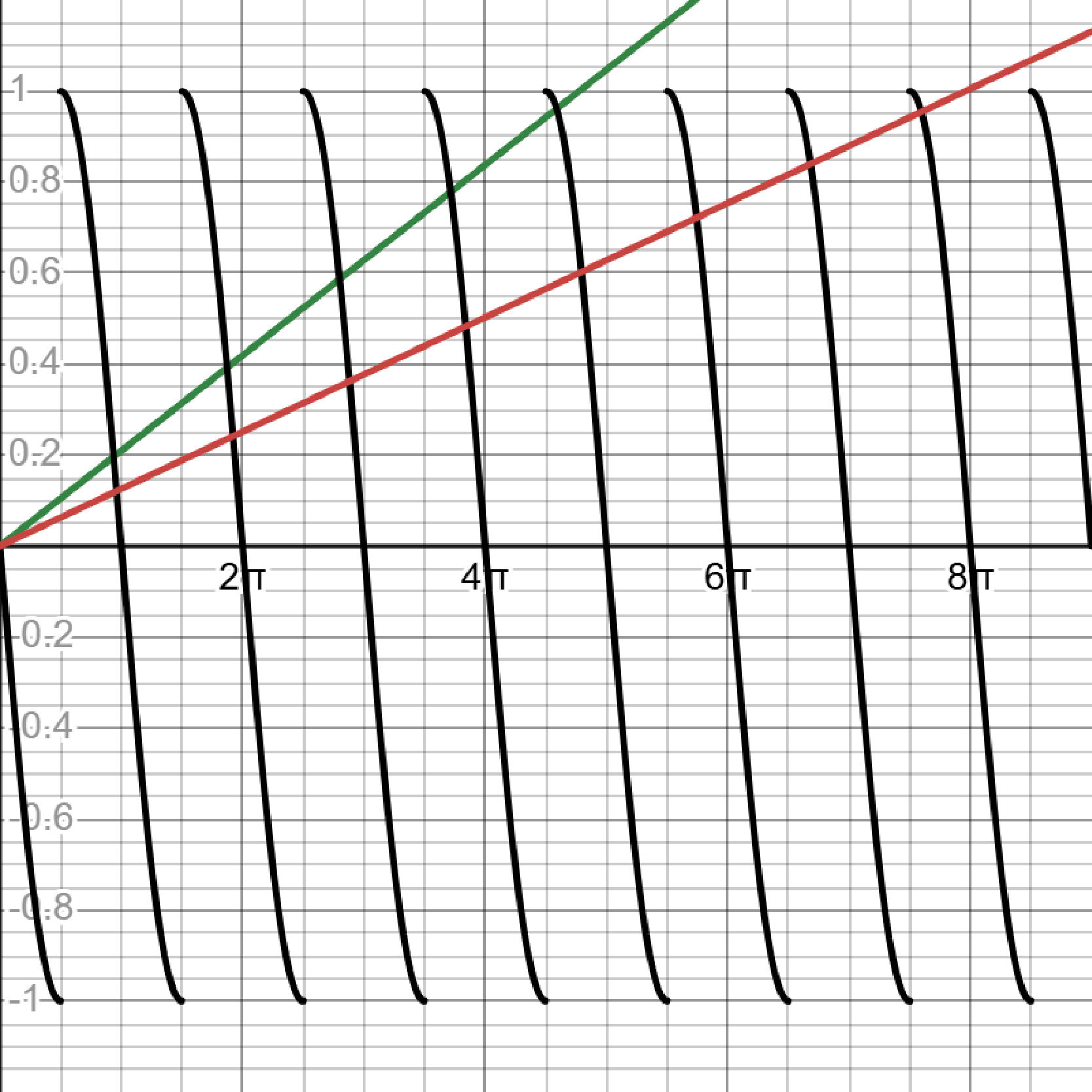}
\caption{Graphical solutions of equation \eqref{transcendental-sinz-modulus-cosz} for $z_0=15$ (steeper straight line) and $z_0=25$ (less slanted straight line). }
\label{Figure-intersections-z-sinz-modulus-cosz}
\end{center}
\end{figure}

Figure \ref{Figure-intersections-z-sinz-modulus-cosz} shows the graphical solutions to equation \eqref{transcendental-sinz-modulus-cosz} for $z_0=15$ and $z_0=25$. The number of intersections and the corresponding values of $z$ are the same as those displayed in Fig. \ref{Figure-intersections}.    The discontinuities of the right-hand side of equation \eqref{transcendental-sinz-modulus-cosz} at $z = m\pi/2$, $m$ odd, naturally rule out the intersections at such $z$-values as valid solutions. As it turns out, however, the  ``simplified'' form \eqref{transcendental-sinz-modulus-cosz} of the transcendental equation \eqref{transcendental-equation-exact} is not so much simpler after all. For the purpose of achieving graphical solutions, it may be preferable to stick to equation \eqref{transcendental-equation-exact}. Nevertheless, equation \eqref{transcendental-sinz-modulus-cosz} turns out to be very expedient for finding extremely accurate  approximations to the energy levels, as will be shown in the next section.

If $z_0$ is very large, the straight line ${\tilde z}=z/z_0$ is nearly coincident with the $z$-axis and Fig. \ref{Figure-intersections-z-sinz-modulus-cosz} shows that the intersections occur almost exactly at $z_n=n\pi$, where $n$ is a positive integer. Thus, as expected, in the limit $V_0 \to \infty$, which implies $z_0 \to \infty$ by  \eqref{z-ztilde-z-zero}, the system reduces to the particle in a box, which  according  to equation \eqref{allowed-energy} has the well-known energy levels  $\displaystyle E_n =\frac{n^2\pi^2\hbar^2}{2ma^2}$. 

\section{Approximate energy spectrum}\label{Newton-method}

Newton's method is a powerful iterative scheme for finding zeroes of functions. The strength  of the  method lies in its extraordinarily rapid rate of convergence \cite{Courant,Spivak}. Suppose one is searching for a solution to $f(z)=0$ in a certain interval $(a,b)$. After guessing an initial approximate solution $z^{(0)} \in (a,b)$, under propitious conditions Newton's method engenders a convergent sequence $z^{(1)}, z^{(2)}, \dots$ of ever better approximations to the true solution which are iteratively given by 
\begin{equation}
\label{Newton}
z^{(n+1)} = z^{(n)} - \frac{f( z^{(n)} )}{f^{\prime}( z^{(n)} )}.
\end{equation}

Owing to \eqref{transcendental-sinz-modulus-cosz}, the relevant function whose zeroes are to be found  is
\begin{equation}
\label{f(z)-Newton}
f(z) = z + z_0 \frac{\sin z \cos z}{\vert \cos z\vert}.
\end{equation}
As we already know, the successive zeroes of this function belong to the intervals
\begin{equation}
\label{f(z)-zeroes-intervals}
(2m-1)\frac{\pi}{2} < z < m \pi, \qquad m=1,2,3, \ldots 
\end{equation}
with $z$ in the second quadrant (for $m$  odd) or in the fourth quadrant (for $m$ even). Because $\cos z <0$ if $m$ is odd whereas $\cos z > 0$ if $m$ is even, the appropriate function for the $m$-th interval is
\begin{equation}
\label{f(z)-Newton-mth-interval}
f(z) = z + (-1)^mz_0 \sin z.
\end{equation}
Therefore, the iterative procedure \eqref{Newton} takes the form
\begin{equation}
\label{Newton-mth-solution}
z^{(n+1)} = z^{(n)} - \frac{z^{(n)} + (-1)^mz_0 \sin z^{(n)}}{1 + (-1)^mz_0 \cos z^{(n)}}.
\end{equation}
Note that $f^{\prime}(z) = 1 + (-1)^mz_0 \cos z >0$ and $f^{\prime\prime}(z) =  - (-1)^mz_0 \sin z >0$ regardless of whether $m$ is odd or even. These are favorable conditions for the success of Newton's method \cite{Courant,Spivak}. For each $m$, a reasonable initial guess is the midpoint of the corresponding interval, namely
\begin{equation}
\label{initial-guess}
z^{(0)} = (2m-1)\frac{\pi}{2} + \frac{\pi}{4} = (4m-1) \frac{\pi}{4}.
\end{equation}

{\bf Example 1.}  Ground state in the case $z_0=15$.  We have $m=1$, $z^{(0)}=  3\pi/4 \approx 2.35619449$ and the iterative scheme
\begin{equation}
\label{Newton-z=25-m=4}
z^{(n+1)} = z^{(n)} - \frac{z^{(n)} -15 \sin z^{(n)}}{1 - 15 \cos z^{(n)}}.
\end{equation}
The first three iterations give
\begin{equation}
\label{m=1-four-iterations}
z^{(1)} \approx 3.0670319 ; \qquad z^{(2)} \approx 2.9448601 ; \qquad  z^{(3)} \approx 2.9440409.
\end{equation}
According to the results in Section \ref{graphical}, the second iterate is correct to three decimals and the third iterate is correct to six decimals.

\medskip

{\bf Example 2.} Third excited state (fourth energy level)  in the case $z_0=25$. We have $m=4$, $z^{(0)}= 15\pi/4 \approx 11.78097245$ and the iterative scheme
\begin{equation}
\label{Newton-z=15-m=1}
z^{(n+1)} = z^{(n)} - \frac{z^{(n)} + 25 \sin z^{(n)}}{1 + 25 \cos z^{(n)}}.
\end{equation}
The first three iterations give
\begin{equation}
\label{m=4-three-iterations}
z^{(1)} \approx 12.0966808 ; \qquad z^{(2)} \approx 12.0631322; \qquad  z^{(3)} \approx 12.0628480.
\end{equation}
Upon rounding the numbers, the results  in Section \ref{graphical} show that  the second iterate is correct to three decimals and the third iterate is correct to six decimals. 

\medskip

The convergence is very fast in both examples. The number of correct decimals is typically doubled at each iteration \cite{Courant,Spivak}. The method is highly efficient  for finding the energy not only of the ground state but also of the excited states.

\section{A class of exact solutions}\label{Exact}

For certain values of $V_0$ there are exact solutions to \eqref{transcendental-equation-exact} or \eqref{transcendental-sinz-modulus-cosz} 
from which one can find the associated wave function. 

By way of example, suppose
\begin{equation}
\label{z-exact-solutions}
z = 2n\pi + \frac{3\pi}{4} = (8n+3)\frac{\pi}{4}, \qquad n=0, 1, 2, \ldots
\end{equation}
are solutions to \eqref{transcendental-sinz-modulus-cosz}. Then, since $\cos z < 0$ and $\sin z =\sqrt{2}/2$, 
\begin{equation}
\label{z-zero-exact-solutions}
z_0 = \sqrt{2}(8n+3)\frac{\pi}{4}
\end{equation}
and also
\begin{equation}
\label{z-tilde-exact-solutions}
{\tilde z} = \sqrt{z_0^2-z^2} = (8n+3)\frac{\pi}{4} = z.
\end{equation}
As a consequence, from \eqref{z-ztilde-z-zero},
\begin{equation}
\label{V-zero-exact-solutions}
V_0 =  \frac{(8n+3)^2\pi^2\hbar^2}{16ma^2} 
\end{equation}
whereas, from \eqref{allowed-energy},
\begin{equation}
\label{energy-exact-solutions}
E =  \frac{(8n+3)^2\pi^2\hbar^2}{32ma^2} = \frac{V_0}{2}.
\end{equation}
In order to avoid misunderstandings, let us emphasize that these energies {\it are not} energy levels associated with a {\it single} well depth  $V_0$. For each $n$, equation \eqref{energy-exact-solutions} furnishes a {\it single} allowed energy for the {\it especific} $V_0$ given by equation \eqref{V-zero-exact-solutions}.  

And now for the wave function associated with the above energy. From equation \eqref{boundary-conditions-x=a-1} it follows that 
\begin{equation}
\label{B-in-terms-A}
B = e^{{\tilde k}a} A \sin z = e^{ka} \frac{A}{\sqrt{2}},
\end{equation}
where we have used ${\tilde k} = k$. Therefore, the wave function is
 \begin{equation}
\label{wave-function}
 \psi (x) = A\left\{ \begin{array}{cl}
                       \sin kx  & \mbox{if $0 \leq x \leq a$}\\
                     \displaystyle \frac{e^{-k(x-a)}}{\sqrt{2}} & \mbox{if $x>a$}
                  \end{array} \right.
\end{equation}
where the normalization constant $A$ can be taken as real and positive --- of course the negative portion of the $x$-axis has been disregarded because $\psi =0$ there. Normalization requires
\begin{equation}
\label{normalization}
A^{-2} = \int_0^a \sin^2 kx \, dx + \frac{1}{2}\int_a^{\infty} e^{-2k(x-a)}\, dx \equiv I_1 + I_2.
\end{equation}
 Taking into account that $k = z/a$, the first of the above integrals is easily computed as
\begin{equation}
\label{first-integral}
I_1 = \int_0^a \sin^2 \frac{zx}{a}\, dx  = \frac{a}{z}\int_0^z \sin^2 u \, du = \frac{a}{z}\frac{2z-\sin 2z}{4} = \frac{a}{z}\frac{2z+1}{4},
\end{equation}
where we have used
\begin{equation}
\label{sin-2z}
\sin 2z = \sin \bigg(4n\pi + \frac{3\pi}{2}\bigg) = \sin \frac{3\pi}{2} = -1.
\end{equation}
As to the second integral in equation \eqref{normalization}, we have
\begin{equation}
\label{second-integral}
I_2 = \frac{1}{2}\int_a^{\infty} e^{-2k(x-a)}\, dx = \frac{1}{2}\int_0^{\infty} e^{-2kx}\, dx = \frac{1}{2} \frac{1}{2k} = \frac{a}{4z}.
\end{equation}
Consequently, equation \eqref{normalization} yields 
\begin{equation}
\label{normalization2}
A^{-2} = I_1 + I_2 = \frac{a}{2}\frac{z+1}{z} \quad \Longrightarrow \quad A = \sqrt{\frac{(8n+3)\pi}{(8n+3)\pi + 4}\, \frac{2}{a}},
\end{equation}
where \eqref{z-exact-solutions} has been used.

The probability of finding the particle inside the well is
\begin{equation}
\label{probability}
P_n  =  \int_0^a \vert \psi (x)\vert^2\, dx = A^2 \int_0^a \sin^2 kx \, dx = A^2 I_1=
 \frac{2}{a}\frac{z}{z+1}\, \frac{a}{z}\frac{2z+1}{4}   = \frac{4z+2}{4z+4},
\end{equation}
where equations \eqref{first-integral} and \eqref{normalization2} have been used. With the help of equation \eqref{z-exact-solutions} one finally gets
\begin{equation}
\label{probability-final}
P_n = \frac{(8n+3)\pi +2}{(8n+3)\pi +4}.
\end{equation}
Here are a few values of this probability:
\begin{equation}
\label{probability-values}
P_0 \approx 0.851=85.1 \%; \qquad P_{10} \approx 0.992=99.2 \%;\qquad P_{100} \approx 0.999 =99.9 \%.
\end{equation}

Although the value of the energy is always half the well depth, the probability of finding the particle within the well grows steadily as $n$ increases. This probability is one in the limit $n \to \infty$, which is expected because by equation \eqref{V-zero-exact-solutions} this is exactly the limit of an infinitely deep well (particle in a box). One might think that the probability of finding the particle inside the well should  be independent of $n$ because it should depend only on the ratio $E/V_0$, which is $1/2$ for all $n$. This seems to be supported by the fact that the reflection and transmission coefficients for a step potential of height $V_0$ depend only on the ratio $E/V_0$ if the energy is larger than $V_0$ \cite{Eisberg}. But the step potential  problem contains no  length scale and the reflection and transmission coefficients cannot  depend even on the Planck constant because they are dimensionless magnitudes that can depend only on $E/V_0$, which is the sole dimensionless quantity that can be constructed. Dissimilarly, the probability of tunneling across a barrier of width $a$ depends  not only on the ratio $E/V_0$ but also on $V_0$, $a$, $\hbar$ and the particle's mass \cite{Eisberg}. This is because $V_0$, $a$, $\hbar$ and $m$  can be combined into the dimensionless quantity $mV_0a^2/\hbar^2$, and now there are two independent dimensionless quantities.  The same behavior is expected in the present situation involving a potential well whose width  $a$ introduces a length scale in the problem. Indeed, if $V_0 \to \infty$ then, in virtue of equation \eqref{k-tilde}, it is also the case that ${\tilde k} \to \infty$ even if  $E$ is arbitrarily  close to $V_0$ while remaining less than $V_0$. As a consequence, equation \eqref{psi-outside} implies that the wave function vanishes outside the wall in this limit.

\section{Conclusion}\label{Conclusion}

The semi-infinite potential well is an interesting theoretical model with practical  applications to realistic physical systems. The allowed energies are determined by a transcendental equation, which has been graphically solved. Inequalities have been provided by means of which one easily finds the number of bound states in terms of  the well's depth.  The aforesaid transcendental equation can be simplified with the use of simple trigonometric identities. But we have shown that this must be done carefully lest false solutions be created and true solutions be missed. The simplified transcendental equation turns out to be  particularly suited for obtaining remarkably accurate approximations to the energy levels. 
We have also exhibited a class of exact solutions for the energy whose associated wave functions have been exactly normalized. For these stationary states the exact probability of finding the particle inside the well has been calculated. All things considered, this  is a nice problem for a modern physics or introductory quantum mechanics course.  




\end{document}